\documentclass[pra,singlecolumn,preprint,footinbib,amssymb,longbibliography,amsfonts,amsmath,superscriptaddress,showpacs,hyperref]{revtex4-1}
\usepackage{graphicx}
\usepackage{bm}% bold math
\usepackage{color}
\usepackage{url}
\usepackage{epstopdf}
\usepackage{amsthm,mathrsfs} %mathematical symbols
\usepackage{txfonts}
\usepackage[colorlinks=true,linkcolor=blue,urlcolor=blue,citecolor=blue,pdfusetitle]{hyperref}
\usepackage{soul}
\usepackage{lineno}
\newcommand{\xb}{\mathbf{x}}

\begin{document}
\title{Entropy Production in Continuously Measured Gaussian Quantum Systems}

\author{Alessio Belenchia}
\affiliation{Centre for Theoretical Atomic, Molecular, and Optical Physics, School of Mathematics and Physics, Queens University, Belfast BT7 1NN, United Kingdom}
\author{Luca Mancino}
\affiliation{Centre for Theoretical Atomic, Molecular, and Optical Physics, School of Mathematics and Physics, Queens University, Belfast BT7 1NN, United Kingdom}
\author{Gabriel T. Landi}
\affiliation{Instituto de F\'isica, Universidade de S\~{a}o Paulo, CEP 05314-970, S\~{a}o Paulo, S\~{a}o Paulo, Brazil}
\author{Mauro Paternostro}
\email{m.paternostro@qub.ac.uk}
\affiliation{Centre for Theoretical Atomic, Molecular, and Optical Physics, School of Mathematics and Physics, Queens University, Belfast BT7 1NN, United Kingdom}

\date{\today}

\begin{abstract}
The entropy production rate is a key quantity in non-equilibrium thermodynamics of both classical and quantum processes. No universal theory of entropy production is available to date, which hinders progress towards its full grasping. By using a phase space-based approach, here we take the current framework for the assessment of thermodynamic irreversibility all the way to quantum regimes by characterizing entropy production -- and its rate -- resulting from the continuous monitoring of  a Gaussian system. This allows us to formulate a sharpened  second law of thermodynamics that accounts for the measurement back-action and information gain from a continuously monitored system. We illustrate our framework in a series of physically relevant examples. 
\end{abstract}
\maketitle

%----------------------------------------------------------------------------------------------------------------------------

\begin{flushleft}
\section*{Introduction}

 Entropy production, a fundamental concept in non-equilibrium thermodynamics, provides a measure of the degree of irreversibility of a physical process. It is of paramount importance for the characterization of an ample range of systems across all scales, from macroscopic to microscopic~\cite{Onsager1931, Machlup1953, deGroot1961,Tisza1957, Schnakenberg1976, Tome2012, Landi2013, Breuer2003,  Deffner2011, deOliveira2016, Batalhao2019}. The lack of a continuity equation for entropy prevents entropy production from being a physical observable, in general. Its quantification must thus pass through inference strategies that connect the values taken by such quantity to accessible observables, such as energy~\cite{Crooks1998,Jarzynski1997,Jarzynski2004}. This approach has recently led to the possibility to experimentally measure entropy production in microscopic~\cite{Batalhao2015} and mesoscopic quantum systems~\cite{Brunelli2018}, and opened up intriguing opportunities for its control~\cite{Micadei2019}. Alternative approaches to the quantification of entropy production are based on the ratio between forward and time-reversed path probabilities of trajectories followed by systems undergoing non-equilibrium processes~\cite{Crooks2000,Spinney2012,Manzano2018}. 

A large body of work has been produced in an attempt to overcome the lack of generally applicable theories of entropy production~\cite{Batalhao2019,Landi2019}. Remarkably, this has allowed the identification of important contributions to the irreversibility emerging from a given process stemming from system-environment correlations~\cite{Esposito2010,Reeb2014,Micadei2019}, quantum coherence~\cite{Santos2019,Francica2019,Mohammady2019}, and the finite size of the environment~\cite{Reeb2014,Santos2019rapid}.

Among the directions of potential further investigations, {a particularly relevant one} is the inclusion of the back-action resulting from measuring a quantum system. Measurements {can} have a dramatic effect on both the state and the ensuing dynamics of a quantum system: while the randomness brought about by a quantum measurement adds stochasticity to the evolution of a system, the information gained through a measurement process unlocks effects akin to those of a Maxwell daemon~\cite{Koski2015,Elouard2017}. Both such features are intuitively expected to affect the entropy production and its rate [cf. Fig.~\ref{ClaLim}] 

\begin{figure}[t!]
\centering
\includegraphics[width=\columnwidth]{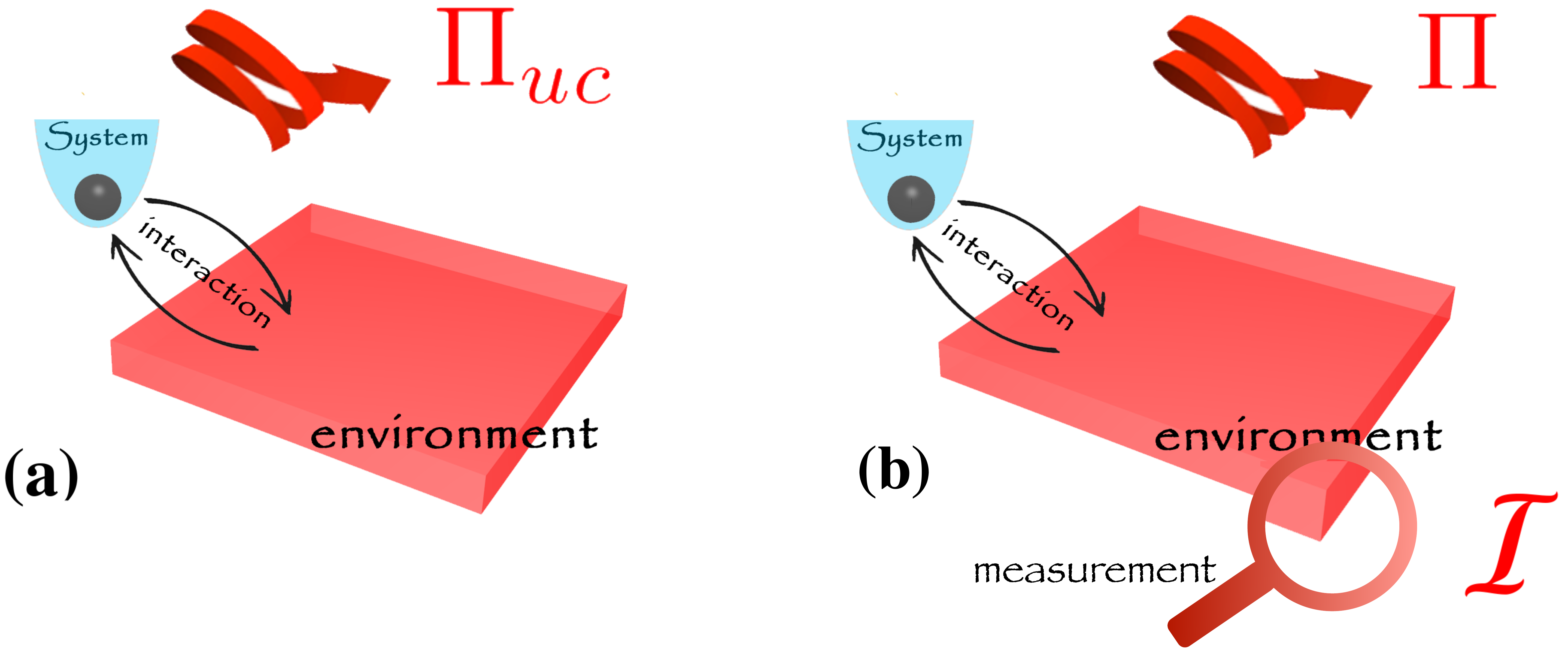}
  \caption{{\bf Scheme of principle}. {\bf (a)} A system, prepared in an arbitrary state and being externally driven, interacts with an environment. The dynamics is associated with a (unconditional) rate of entropy production $\Pi_{\rm{uc}}$. {\bf (b)} Continuous measurements alter the dynamics of the system, resulting in an entropy production rate $\Pi$ that includes an information-theoretical contribution $\dot{{\cal I}}$ determined by the amount of information extracted through the measurement.}
   \label{ClaLim}
\end{figure}

The ways such modifications occur have been the focus of some attention recently. Ref.~\cite{Elouard2017,PhysRevA.99.022117} tackled the problem by focusing on the stochastic energy fluctuations that occur during measurements, while Refs.~\cite{Alonso2016,elouard2017role,Stefano2018, Naghiloo2017,Naghiloo2018,Horowitz2012,Hekking2013,cottet2017observing,masuyama2018information} addressed the case of weak quantum measurements of a system, which allowed for the introduction of trajectory-dependent work and heat quantifiers (cf. Refs.~\cite{strasberg2019stochastic,strasberg2019repeated}). In line with a Landauer-like framework, Sagawa and Ueda focused on the minimum thermodynamic cost implied by a measurement, highlighting the information theoretical implications of the latter on the stochastic thermodynamics of a two-level system~\cite{Sagawa2008,Sagawa2009}, an approach that can both be generalized to general quantum measurements~\cite{Abdelkhalek2016} and assessed experimentally~\cite{mancino2018entropic}.

Here we contribute to the quest for a general framework for entropy production and its rate in a general system brought out of equilibrium and being continuously monitored. We propose a widely applicable formalism able to identify the measurement-affected rate of entropy production and the associated entropy flux to or from the environment that is connected to the monitored system. We unveil the thermodynamic consequences of measuring by way of both general arguments and specific case {studies}, showing that the entropy production rate can be split in a term that is intrinsically dynamical and one that is informational in nature. We show that, for continuously measured Gaussian systems, the entropy production splitting leads to a refined, tighter  second law. We also discuss the possibility to control the non-equilibrium thermodynamics of a system through suitable measurements. We illustrate it by studying a thermal quench of an harmonic oscillator and a driven-dissipative optical parametric oscillator. 

Our approach paves the way to the assessment of the non-equilibrium thermodynamics of continuously monitored Gaussian systems -- such as (ensembles of) trapped ions and quadratically-confined levitated optomechanical systems -- whose energetics will require tools designed to tackle the intricacies of quantum dynamics and the stochasticity of quantum measurements, which are currently lacking~\cite{Genoni2015,GenoniRuss,Millen2014,Gieseler2014,Vinante2019,Debiossac2019,Rondin2017,Ricci2017,Gieseler2018,Hoang2018}.

\section*{Results}
\noindent
{\emph{Entropy rate of a continuously measured system.--} 
The dynamics of a continuously measured Markovian open quantum system can be described by a Stochastic Master Equation (SME) that describes the evolution \textit{conditioned} on the outcomes of the continuous measurement.~\cite{Doherty1999,serafini2017quantum,Genoni2016}. Upon averaging over all trajectories, weighted by the outcomes probabilities, the stochastic part vanishes leaving a deterministic Lindblad ME for the system, whose dynamics we call \textit{unconditional}. } 

{
In the unconditional case, the entropy rate can be split as $\dot{S}_{\rm{uc}}=\Phi_{\rm{uc}}+\Pi_{\rm{uc}}$ with $\Phi_{\rm{uc}}$ ($\Pi_{\rm{uc}}$) the unconditional entropy flux (production) rate~\cite{Batalhao2019}, with $S_{\rm{uc}}$ \textit{an} entropic measure. The $2^{nd}$ law of thermodynamics for the unconditional dynamics is encoded in $\Pi_{{\rm uc}}\geq 0$. It should be noted that, choosing the widely used von Neumann entropy leads to controversial results when the system is in contact with a vanishing-temperature thermal bath~\cite{Santos2018,uzdin2018second}. We will address this point again later on.
}

{
For the conditional dynamics a similar splitting of the entropy rate can be obtained, although involving stochastic trajectory-dependent quantities. Indeed, given that the conditional dynamics describes quantum trajectories in Hilbert space, the entropy rate and the entropy production are stochastic quantities. We are interested in the average of such stochastic quantities over all trajectories, i.e. 
\begin{equation}\label{EntrSplit}
    \mathbb{E}[dS/dt]=\mathbb{E}\left[d\phi/dt\right]+\mathbb{E}\left[ d\pi/dt\right]= \Phi+\Pi,
\end{equation} 
where $dS,\,d\pi,\,d\phi$ depend on the stochastic conditional state, and $\Pi$ and $\Phi$ are the averaged conditional entropy production and flux rates. {These quantities will in general differ from their unconditional counterparts because they depend non-linearly on the state of the system. We now move on to explain in more detail their physical interpretation.} %\textcolor{red}{The physical meaning of such quantities will be clarified below.}
}

{Operationally, the conditional dynamics is obtained by continuously monitoring the system and recording the outcomes of the measurement. After repeating the experiments many times, and computing each time the stochastic entropy flux and entropy production rate, one finally averages these quantities over the stochastic trajectories of the system. Such average quantities are related to the state of the system conditioned on the outcomes of the continuous measurements. On the contrary, the unconditional dynamics is recovered by ignoring the outcomes of the measurements at the level of the state of the system. Thus, the \textit{state} of the system, from which the statistics are inferred, corresponds to the average over the possible outcomes of the conditional state at each instant of time.}

{
We aim to connect $\Pi$ to the entropy produced by the system due to only the {open system} dynamics ($\Pi_{{\rm uc}}$). We thus \textit{assume} that the stochastic entropy flux is linear in the conditioned state of the system, i.e., $\mathbb{E}[d\phi/dt]=\Phi_{\rm{uc}}$. This is so in all those cases where the entropy flux is given by the heat flux from the system to an (equilibrium) environment at a reference temperature, a second relevant instance being when the entropy flux is associated to the occurrence of a quantum jump of the system~\cite{Breuer2003}. To the best of our knowledge, no example of violation of this assumption has been re- ported in the literature so far. However, a formal proof of linearity is still missing. %In that case, the heat flux rate is by definition linear in the conditioned state. 
As we aim at a general framework that encompasses non-equilibrium states of the environment, here we  %go beyond the assumption of equilibrium environments and do 
do not necessarily identify the entropy flux rate with the heat flow rate to a thermal environment, while claiming for its linearity in the conditional state of the system. 
{ In what follows we show that, for Gaussian systems, this linearity can be explicitly proven. The same can be done for more general systems starting from a microscopic description of the system-environment interaction via a repeated collisions model~\cite{inpreparation}.}
% Another argument in support of the linearity of the entropy flux on the state of the system is in Ref.~\cite{Breuer2003}, where t 
}

{By comparing the splitting of the entropy rate for conditioned and unconditioned dynamics, we arrive at 
\begin{equation}\label{irreEn}
\Pi=\Pi_{{\rm uc}}+\dot{\mathcal{I}},
\end{equation}
where the last term quantifies the entropic cost of continuously monitoring the system and, {as $\Phi=\Phi_{\rm{uc}}$, it must then follow  $\dot{\mathcal{I}}=\mathbb{E}[dS/dt]-dS_{{\rm uc}}/dt$.} %By adopting the von Neumann entropy as a quantifier, it is straightforward to check that 
%\begin{equation}\label{relativeEn}
%\mathbb{E}[S]-S_{{\rm uc}}=-\sum_{J} p_J S(\rho_J||\rho_{{\rm uc}}),
%\end{equation}    
%where $J$ stands for the stochastic trajectories of the system, $p_J$ and $\rho_J$ for their respective probabilities and conditioned quantum state, and $S(\rho||\sigma)$ is the quantum relative entropy. 
}

In order to further characterize the \textit{informational} term $\dot{\mathcal{I}}$ and, in particular, single out the entropy production and flux rates, we focus on open Gaussian systems subject to continuous Gaussian measurements. Such a class of systems and processes plays a substantive role in the broad panorama of quantum optics, condensed matter physics, and quantum information science in general. Gaussian measurements are some of the most widely used techniques in quantum labs, and their role in stochastic non-classical thermodynamics is thus both interesting and physically very well motivated.

The Gaussian case calls for the use of powerful phase-space techniques, in conjunction with the adoption of the Wigner entropy as an entropic measure~\cite{Santos2018}, which allow us to unambiguously identify the entropy flux and production rates based on the dynamics of the system, without resorting to a thermal environment (at finite temperature) and  bypasses some of the controversies linked to more standard von Neumann entropy-based formulations~\cite{Santos2018}, thus going beyond any standard approach. %The Wigner entropy, henceforth denoted by $S$,  was recently used to build a quantifier for the entropy production rate of an unconditional Gaussian system that bypasses some of the controversies linked to more standard von Neumann entropy-based formulations~\cite{Santos2018}. 
It should be noted that, in some cases of continuously measured quantum Gaussian systems, e.g., homodyne detection of an optical mode where the bath mode being monitored is effectively at zero temperature, there are no alternatives to the adoption of the Wigner entropy due to the unphysical divergences that plague the definition of entropy production and fluxes obtained by adopting the von Neumann entropy~\cite{Santos2018,uzdin2018second}. Nonetheless, consistently with classical stochastic thermodynamics, Ref.~\cite{Santos2018} showed that the results obtained using the Wigner entropy agree with the von Neumann ones, and crucially with the classical ones, for systems interacting with thermal baths in the high-temperature limit. Finally, we would like to remark that Refs.~\cite{SantosSpin,PhysRevResearch.2.013136} made a step forward to extend phase space-based formulation of entropy production rate and flux to the case of non-Gaussian systems and dynamics, using the well-known Wehrl entropy defined in terms of the Shannon entropy of the Husimi function. All the considerations reported here can be generalised straightforwardly  to the use of the Wehrl entropy.

\noindent
\emph{Continuously measured Gaussian systems.--} Solving the SME is in general a tall order. Luckily, the intricacy of such an approach is greatly simplified when dealing with Gaussian systems, as their description can be reduced to the knowledge of the first two statistical moments of the quadratures of the system. The SME can thus be superseded by a simpler system of stochastic equations. We consider a system of $n$ modes, each described by quadrature operators  $(\hat{q}_i,\hat{p}_i)$ with $[\hat{q}_{j},\hat{p}_j]=i$, and define the vector $\hat{\xb}=(\hat{q}_1,\hat{p}_1,\hat{q}_2,\hat{p}_2,\dots,\hat{q}_{n},\hat{p}_{n})$. When restricting to Gaussian systems, the Hamiltonian is at most quadratic in the quadrature operators and can be written as 
%\begin{equation}
    $\hat{H}=\frac{1}{2}\hat{\xb}^{T}H_{s}\hat{\xb}+\mathbf{b}^{T}\Omega\hat{\xb}$,
%\end{equation}
where $H_{s}$ is a $2n\times 2n$ matrix, $\mathbf{b}$ is a $2n-$dimensional vector accounting for a (time-dependent) linear driving, and $\Omega=\oplus^n_{j=1}i\sigma_{y,j}$ is the $n$-mode symplectic matrix ($\sigma_{y,j}$ is the $y$-Pauli matrix of subsystem $j$). %defined as 
For an environment modelled by Lindblad generators that are linear in the quadratures of the system and the latter is monitored through Gaussian measurements, the dynamics preserves the Gaussianity of any initial state. In this case, the vector of average moments $\bar{\xb}=\langle \hat{\xb} \rangle$ and the Covariance Matrix (CM) $\sigma_{ij}=\langle\{\hat{\xb}_{i},\hat{\xb}_{j}\}\rangle/2-\langle \hat{\xb}_{i}\rangle\langle \hat{\xb}_{j}\rangle$ of the modes completely describe the dynamics via the equations~\cite{PhysRevLett.94.070405,Genoni2016,serafini2017quantum}
%If the system is prepared initially in a Gaussian state, the environment is modelled by \textcolor{red}{Lindblad} generators at most linear in the quadratures of the system, and it is monitored via Gaussian measurements, the SME preserves the Gaussianity and the state is characterized by the first momenta $\bar{\xb}=\langle\hat{\xb}\rangle$ and the covariance matrix (CM) $\sigma_{ij}=\langle\{\hat{\xb}_{i},\hat{\xb}_{j}\}\rangle/2-\langle \hat{\xb}_{i}\rangle\langle \hat{\xb}_{j}\rangle$. The dynamical equations for first and second momenta are given by~\cite{}
\begin{equation}
\label{RiccatiEquation}
\begin{aligned}
\dot{\sigma}&=A \sigma + \sigma A^T + D - \chi(\sigma), \\
d\bar{\xb}&=(A\bar{\xb}+\mathbf{b})dt+(\sigma C^T+\Gamma^T)d\mathbf{w},
\end{aligned}
\end{equation}
where $d\mathbf{w}$ is a $2\ell-$dimensional vector of Wiener increments ($\ell$ is the number of output degrees of freedom being monitored), $A$ $(D)$ is the drift (diffusion) matrix characterizing the unconditional open dynamics of the system, and $\chi(\sigma)=(\sigma C^T + \Gamma^T)(C\sigma+\Gamma)\ge0$ is defined in terms of the $2\ell\times 2n$ matrices $C$ and $\Gamma$ that describe the measurement process {(see Ref.~\cite{serafini2017quantum} and references therein for a detailed derivation of these equations)}. While the explicit form of such matrices is inessential for our scopes (cf. Ref.~\cite{PhysRevLett.94.070405,Genoni2016,serafini2017quantum}), we highlight the fact that the drift matrix $A$ can be decomposed as $A=\Omega H_{s}+A_{\rm{irr}}$, where the first term accounts for the unitary evolution and the second for diffusion. Notwithstanding the stochasticity of the overall dynamics, the equation for the covariance matrix is deterministic. Thus, $\sigma(t)$ does not depend on the explicit outcomes of the measurement (i.e., the trajectory followed by the system), while it depends on the measurement carried out~\cite{Genoni2016}. The dynamics of the corresponding unconditional quantities $\sigma_{\rm{uc}}$ and $\bar{\xb}_{\rm{uc}}$ is achieved from Eqs.~\eqref{RiccatiEquation} by taking $C=\Gamma=\mathbb{O}$ with $\mathbb{O}$ the null $2\ell\times 2n$ matrix. {This then implies $\chi(\sigma) = 0$, so that one recovers the dynamical equation for the evolution of $\sigma_\text{uc}$.}

\noindent
\emph{Entropy production rate and flux.--}  Eqs.~\eqref{RiccatiEquation} can be conveniently cast in the phase-space as the continuity equation [cf. Supplementary Information]
\begin{align}\label{StFP}
    dW&=-{\rm div}[J dt+J_{\rm{sto}}],
\end{align}
where $W=e^{-\frac{1}{2}(\xb-\bar{\xb})^T\sigma^{-1}(\xb-\bar{\xb})}/{(2\pi)^n\sqrt{\det\sigma}}$ is the Wigner function associated with the state of the $n$-mode system and we have introduced the deterministic phase-space current $J$, which can be divided as $J=J_{\rm rev}+J_{\rm irr}$, and its stochastic counterpart $J_{\rm{sto}}$. Here, $J_{\rm{rev}}=\Omega H_s \xb W+\mathbf{b} W$ encodes the contribution stemming from the unitary dynamics and $J_{\rm{irr}}=A_{\rm{irr}}\xb W-({D}/{2})\nabla W$ accounts for the irreversible dissipative evolution. It should be noted that these currents are equal to the ones of the unconditional dynamics with the replacement $W\rightarrow W_{\rm{uc}}$, where $W_{\rm{uc}}$ is the Wigner function of the unconditional state obtained through the replacements $\sigma\to\sigma_{\rm{uc}}$ and $\bar{\xb}\to\bar{\xb}_{\rm{uc}}$.
The stochastic term $J_{\rm{sto}}=W(\sigma C^T+\Gamma^T)d\mathbf{w}$ depends entirely on the conditional dynamics, through $\sigma$ and $W$,  and the measurement strategy being chosen.

In order to characterize the entropy of the $n$-mode system we adopt the Wigner entropy $S=-\int W\ln{W}\,d^{2n}\xb$ as our entropic measure. {For Gaussian systems~\cite{Adesso2012}
\begin{equation}
S=\frac12\ln(\det\sigma)+k_n=-\ln({\cal P})+\tilde{k}_n,
\end{equation}
with $k_n, \tilde{k}_n$  inessential constants that depends only on the number $n$ of modes involved and ${\cal P}$ the purity of the state of the system, which for a Gaussian state reads ${\cal P}=(\det 2\sigma)^{-1/2}$.} This also coincides with the R\'enyi-2 entropy~\cite{Adesso2012} and tends to the von-Neumann entropy in the classical limit of high temperatures. {Let us note that the Renyi-$\alpha$ entropy for a generic $N$-mode Gaussian state is defined as $S_\alpha(\rho)=\frac{1}{\alpha-1}\sum^N_{i=1}\ln f_\alpha(\sigma_i)$, where $f_\alpha(\sigma_i)=\left(\frac{\sigma_i+1}{2}\right)^\alpha-\left(\frac{\sigma_i-1}{2}\right)^\alpha$ and $\sigma_i$ is the $i^\text{th}$ symplectic eigenvalues of the Gaussian state. A ``classical limit'' for such a quantity can be recovered by considering $\sigma_i\gg1$, i.e. the ``high-temperature limit''. 
In these conditions, it is easy to see that the R\'enyi-$\alpha$ entropy reduces to $S_\alpha(\rho)=N \log\alpha/(\alpha-1)-N\log2+\sum^N_{i=1}\log\sigma_i$. We see that, except for a $\alpha$-dependent constant, this matches the von Neumann entropy.}
%This quantity was recently used to build a quantifier for the entropy production rate of an unconditional Gaussian system that bypasses some of the controversies linked to more standard von Neumann entropy-based formulations~\cite{Santos2018}. 

As the Wigner entropy only depends on the CM of the system, its evolution is deterministic even for continuously measured system, a peculiarity of Gaussian systems. The same then holds true for the {\it entropy rate}  given by
\begin{equation}
\label{entropy-rate}
    \frac{dS}{dt}=\frac{1}{2}\frac{d}{dt}(\rm{Tr}[\log\sigma])=\frac{1}{2}\rm{Tr}[2A+\sigma^{-1}(D-\chi(\sigma))].
\end{equation}
In the unconditional case, the entropy rate is $\dot{S}_{\rm{uc}}=\Phi_{\rm{uc}}+\Pi_{\rm{uc}}$~\cite{Batalhao2019} and both the unconditional entropy flux and production rates depend on the irreversible part of the phase-space current {$J_{\rm irr}$.} 
%as
%\begin{equation}
%\Phi_{\rm{uc}} =-2\int d^{2n}\xb J^{T}_{\rm{irr}}D^{-1}A_{\rm{irr}}\xb,\quad \Pi_{\rm{uc}}=2\int \frac{d^{2n}\xb}{W_{\rm{uc}}}J_{\rm{irr}}^TD^{-1}J_{\rm{irr}}.
%\end{equation}
In the Supplementary Information accompanying this paper we report expressions for these quantities written in terms of $A_{\rm{irr}}$, $\sigma_{\rm{uc}}$, and $D$. For a system interacting with a high-temperature thermal environment, $\Phi_{\rm{uc}}$ coincides with the energy flux from the system to the environment. Thus, this formalism generalizes the usual thermodynamic description in a meaningful manner.

For the continuously measured case, while Eq.~\eqref{entropy-rate} presents a deterministic quantity, both the entropy flux and production rate are inherently stochastic as they depend on the first moments of the quadrature operators. {Nonetheless, using Eq.~\eqref{StFP}, it is possible to single out the entropy production  and flux  rates as the quadratic and linear part of $dS/dt$ in the irreversible currents, respectively. 
%\begin{align}\label{dS}
%    &d\phi_{\bar{\xb}}=2\int \frac{d\xb}{W}J_{\rm{\rm{irr}}}^TD^{-1}J_{\rm{\rm{irr}}}dt -2\int \frac{d\xb}{W}J_{2}^T\chi^{-1}J_{2}dt\\
%    &d\pi_{\bar{\xb}}= -2\int d\xb J^{T}_{\rm{irr}}D^{-1}A_{\rm{irr}}\xb dt,
%\end{align}
%where $J_{2}=\frac{1}{2}\chi\nabla W$ and the dependency of the entropy rates on the stochastic first momenta has been made explicit. 
The entropy rate is thus written as $dS=d\phi_{\bar{\xb}} +d\pi_{\bar{\xb}}$ with $d\phi_{\bar{\xb}}, \, d\pi_{\bar{\xb}}$ the conditioned trajectory-dependent entropy flux and production rates. It can be shown that, upon taking the average ($\mathbb{E}$) over the outcomes of the measurements,  $\mathbb{E}\left[d\phi_{\bar{\xb}}/dt\right]=\Phi_{{\rm uc}}$, demonstrating that the linearity of the stochastic entropy flux in the state of the system is a property of Gaussian systems and does not need to be postulated, as shown in the Supplemental Information. Thus, on average, we can rewrite the entropy as  $\dot{S}=\mathbb{E}\left[d\phi_{\bar{\xb}}/dt\right]+\mathbb{E}\left[ d\pi_{\bar{\xb}}/dt\right]= \Phi_{{\rm uc}}+\Pi$.}

As the entropy of the conditioned system is deterministic, we can write $ \dot{S}=\dot{S}_{\!\rm uc}+\dot{\mathcal{I}}$,
where the term $\dot{\mathcal{I}}$ accounts for the excess entropy production resulting from the measurement process.
Indeed, by integrating the above expression, one finds $\mathcal{I}=\ln(\mathcal{P}_{\rm{uc}}/\mathcal{P})$, showing that ${\cal I}$ quantifies the noise to be added to the conditional state to bring it back to its unconditional form. As, in general, we have ${\cal P}\ge{\cal P}_{\rm{uc}}$, we gather that ${\cal I}\le0$.  
In fact, it can be further shown that  
$\mathcal{I}=-I(\mathbf{X}:\mathbf{\bar{X}})\leq 0,$
with $I$ the classical mutual information between the phase-space position $\mathbf{X}$ in the unconditional case and the stochastic first moments $\bar{\mathbf{X}}$, which evolve according to the It\^o equation in Eq.~\eqref{RiccatiEquation}. The inequality is saturated iff $\sigma=\sigma_{\rm{uc}}$ [see Supplementary Information for details]. %Note that, the identification of $\mathcal{I}$ with minus the mutual information between conditional and unconditional dynamics is also consistent with the expression in Eq.~\eqref{relativeEn}.

A straightforward calculation based on Eq.~\eqref{entropy-rate} leads to  
{
\begin{equation}
\label{I}
    \dot{\mathcal{I}}=\frac{1}{2}\rm{Tr}[\sigma^{-1}(D-\chi(\sigma))-\sigma_{\rm{uc}}^{-1}D],
\end{equation}
where, as seen from Eq.~\eqref{RiccatiEquation}, $D-\chi(\sigma)$ accounts for a modification to the diffusion matrix of the dynamics} occurring due to the choice of measurement: the latter conjures with the environment and acts on the system so as to modify its diffusive dynamics. Eq.~\eqref{I} embodies the effects of continuous detection, which modifies the CM of the system and its dynamics. 

{As for Eq.~\eqref{irreEn}, we have $ \Pi=\Pi_{\rm{uc}}+\dot{\mathcal{I}}.$ This is the main result of this work.} It connects the entropy production rate of the unmonitored open Gaussian system to the homonymous quantity for the monitored one via the informational term $\dot{\mathcal{I}}$. The second law for the unmonitored system $\Pi_{\rm{uc}}\geq 0$ can now be used to obtain the refined second law for continuously measured Gaussian systems $\Pi\geq \dot{\mathcal{I}}$, which epitomizes the connection between non-equilibrium thermodynamics and information theory, as pioneered by Landauer's principle. The degree of irreversibility of the dynamics being considered, which is associated with a change in entropy of the state of the system, is lower-bounded by an information theoretical cost rate that depends on the chosen measurement and that is in general more stringent than that associated with the unmeasured dynamics. This echoes and extends significantly the results in Ref.~\cite{Sagawa2009}, which were valid for discrete measurements and equilibrium environments.  

\emph{Case study.--} To illustrate our framework, we consider two relevant examples: a thermal quench and the driven-dissipative optical parametric oscillator. In both cases, the system (either a driven harmonic oscillator or an optical parametric oscillator) is monitored by a single field mode in thermal equilibrium. {We call $\bar{n}_{\rm th}$ the mean occupation number of such thermal state of the field, which is subjected to general-dyne measurements~\cite{serafini2017quantum}.} The system is coupled to the thermal field mode  via an excitation-exchange interaction,
{
\begin{equation}
\hat{H}_{int}=\sqrt{\gamma}\hat{\xb}^T\hat{\xb}_{B},
\end{equation}
where $\hat{\xb}=(\hat{q},\hat{p})$ and $\hat{\xb}_{B}$ represent the quadratures vector for system and bath respectively. Calling $H_s$ the system Hamiltonian in the two cases, the SME describing the conditional dynamics of the continuously monitored system is written as~\cite{serafini2017quantum}
\begin{equation}\label{SME}
\dot{\rho}=-i[H_s,\rho]+\gamma(\bar{n}_{\rm{th}}+1)\mathcal{D}[\hat{a}]\rho+\gamma \bar{n}_{\rm{th}}\mathcal{D}[\hat{a}^\dag]\rho+\sqrt{\gamma}\mathcal{H}[\hat{a}]\rho dw,
\end{equation}
where $\mathcal{D}[\hat{\mathcal{O}}]\rho=\hat{\mathcal{O}}\rho\hat{\mathcal{O}}^\dag-(\hat{\mathcal{O}}^\dag \hat{\mathcal{O}}\rho+\rho \hat{\mathcal{O}}^\dag \hat{\mathcal{O}})/2$, $dw$ is a Wiener increment, and $\mathcal{H}[\hat{\mathcal{O}}]\rho=\hat{\mathcal{O}}\rho+\rho\hat{\mathcal{O}}^\dag-\rm{Tr}\left[\rho(\hat{\mathcal{O}}+\hat{\mathcal{O}}^\dag)\right]\rho$. The last, stochastic term in Eq.~\eqref{SME} encodes the effect of the continuous monitoring of the $\hat{p}$ quadrature of the environment's mode (cf. Ref.~\cite{Genoni2016}). Similar SMEs describe general-dyne measurements (cf. Refs.~\cite{wiseman2001complete,wiseman2009quantum,PhysRevLett.94.070405} for the most general SME for bosonic systems and Markovian dynamics and further details).}

{The two exemplary cases that we consider are I) a driven harmonic oscillator described by $H_s=\omega/2\, (\hat{q}^2+\hat{p}^2)+H_{drive}$, where $H_{drive}=i\mathcal{E}(\hat{a}e^{i\theta}-\hat{a}^{\dag}e^{-i\theta})$ describes the driving by a pump of amplitude $\mathcal{E}$ and phase $\theta$; II) the optical parametric oscillator described by $H_s=-\chi/2\,(\hat{q}\hat{p}+\hat{p}\hat{q})$. From now on we work in natural units and set $\omega=\chi=1$. Further details on these examples and their description in  the Gaussian framework can be found in the Methods.
}

In Fig.~\ref{panels}, the behavior of the entropy production rate, entropy flux rate, and $\dot{\mathcal{I}}$ is shown for both systems under either homodyne or heterodyne measurements. 
{For example I), we have chosen a thermal bath with $2\bar{n}_{\rm th}+1=100$ and a displaced thermal initial state for the system with a larger initial temperature. The system is thus cooled down by the interaction with the environment. For the parametric oscillator in example II), the thermal bath is chosen at zero temperature, the initial state is  a squeezed vacuum state, and we also consider a small additive Gaussian noise so as to make the detection not ideal. Such additional noise makes the steady-state different for different detection strategies.} 

\begin{figure*}[t!]
\centering
\includegraphics[width=0.85\textwidth]{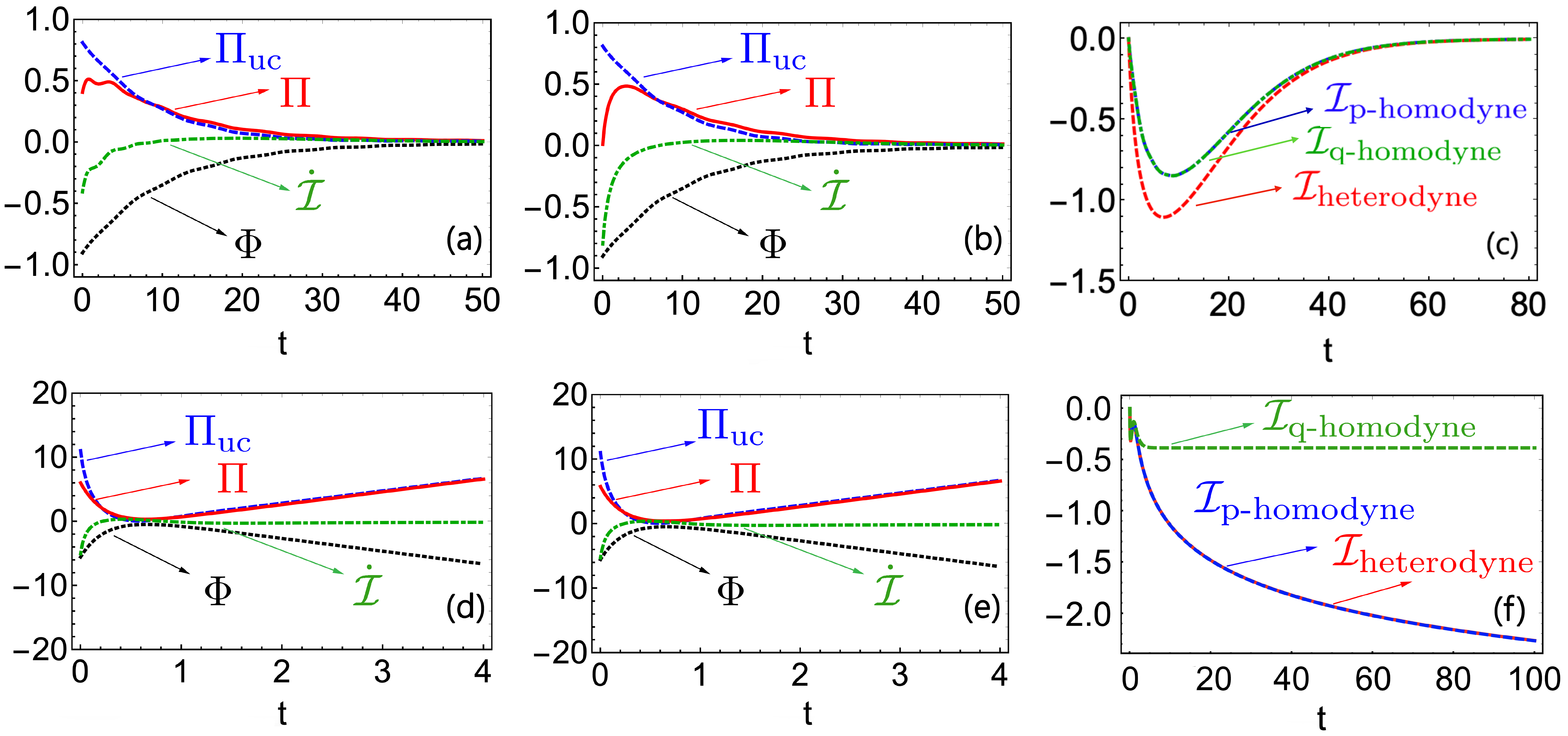}
  \caption{{\bf Numerical example}. {\bf (a)}-{\bf (c)}: Thermal quench of a harmonic oscillator interacting with an environment in thermal equilibrium (mean occupation number {{$\bar{n}_{\rm{th}}$ such that}} $2 \bar{n}_{\rm{th}} + 1 = 100$) and subjected to homodyne and heterodyne measurements. The initial state of the harmonic oscillator, which is externally driven by a pump of amplitude ${\cal E}$, is thermal with an energy ten times larger than the environment. %$\sigma_0=\rm{diag}(\Xi \bar{n}, \Xi \bar{n})$ with $\Xi=10$, and its evolution. 
  Panels {\bf (a)} and {\bf (b)} show $\Pi_{\rm{uc}}$ (dashed blue curve),  $\Pi$ (solid red curve), $\Phi$ (black dotted curve), and $\dot{\mathcal{I}}$ (dot-dashed green curve) for homodyne (monitoring of the system's momentum quadrature) and heterodyne detection. Panel {\bf (c)} shows ${\cal I}$ when homodyning the position and momentum quadratures of the system (dot-dashed green curve and dotted red curve, respectively), and heterodyne measurements (dashed blue curve). {\bf (d)}-{\bf (f)}: Optical parametric oscillator in contact with an external mode in thermal equilibrium ($\bar{n}_{\rm{th}}=0$) and subjected to homodyne and heterodyne measurements with additional Gaussian noise. Panels {\bf (d)} and {\bf (e)} are analogous to {\bf (a)} and {\bf (b)}. Panel {\bf (f)} shows ${\cal I}$ for various detection schemes [cf. Methods].}
   \label{panels}
\end{figure*}

{The behaviour of the  information rate $\dot{\mathcal{I}}$ in both examples is consistent with expectations [cf. Fig.~\ref{panels} {\bf (a)-(b), (d)-(e)}]. It starts at an initially large negative value in light of the non-equilibrium nature of the initial state, but then vanishes as the system approaches the steady-state of the conditional dynamics. In fact, at the steady-state no additional information on the system is acquired with respect to the one acquired up to that moment. This does not mean, however, that the continuous measurement is unnecessary at the steady-state. Indeed, the  continuous monitoring sustains the steady-state, which would otherwise evolve towards its unconditional counterpart effectively causing a loss of information~\cite{rossi2020experimental,inpreparation}.}

{It should be noted that, after an initial negative phase, $\dot{\mathcal{I}}$ becomes positive in both examples, consistently with Fig.~\ref{panels} {\bf(c)-(f)}. While  $\dot{\mathcal{I}}$ takes both positive and negative values, the regions where $\dot{\mathcal{I}}>0$ show how our refined $2^{\rm{nd}}$ law offers a more stringent constraint than the standard one.}

{Finally, by comparing Figs. 2 {\bf(c)} and {\bf(f)}, we note a stark  qualitative difference between the informational terms resulting from the dynamics of a harmonic oscillator subjected to a thermal quench and the optical parametric oscillator. Indeed,} the inspection of the results for the thermal quench shows that $\mathcal{I}\to0$ at the steady-state [cf. Fig.~\ref{panels} {\bf (c)}]. {This is justified by the fact that} the steady-state of the conditional and unconditional dynamics is the same irrespective of the measurement {and asymptotically no extra information from the measurements is acquired with respect to the unconditional case.} In contrast, Fig.~\ref{panels} {\bf (f)}, which refers to the parametric oscillator, shows different long-time behaviour of $\mathcal{I}$ for different measurements due to the additive Gaussian noise that is introduced by the choice of detection strategy. This makes the steady-state of the conditioned dynamics mixed, with a purity dependent on the measurement {and, in turn, the measurement outcomes remain informative also in the asymptotic regime.}

%The inspection of the results for the thermal quench shows that $\mathcal{I}\to0$ at the steady state [cf. Fig.~\ref{panels} {\bf (c)}] as the steady state of conditional and unconditional evolution is the same irrespective of the measurement. 

%In contrast, Fig.~\ref{panels} {\bf (f)}, which refers to the parametric oscillator, shows different long-time behavior of $\mathcal{I}$ for different measurements due to the additive Gaussian noise that is introduced by the choice of detection strategy. This makes the steady state of the conditioned dynamics mixed, with a purity dependent on the measurement. While in both examples $\dot{\mathcal{I}}$ takes both positive and negative values, the regions where $\dot{\mathcal{I}}>0$ showcase how our refined $2^{\rm{nd}}$ law offers a more stringent constraint than the standard one.

\section*{Discussion} We have characterized the entropy production rate of continuously monitored non-equilibrium Gaussian quantum processes in terms of the effects induced by measurement. This led us to the formulation of a refined  $2^{\rm{nd}}$ law reminiscent of Landauer's principle and in line with previous results valid for discrete measurements and systems. {However, it should be  noted that, the particular form of this refined second law is tailored to Gaussian dynamics and the Wigner entropy as an entropic measure}. 

On one hand, our results shine light on the emerging field of information thermodynamics, highlighting the tight fundamental link between non-equilibrium thermodynamics and information gains. On the other hand, they offer a general way to define the entropy production and flux rates for Gaussian systems, thus overcoming the limitations of previous approaches. 

Our framework will be invaluable to analyze and characterize the non-equilibrium dynamics of experimental systems of strong current interest. In particular, levitated quantum optomechanics offers fertile ground for the application of our formalism~\cite{Debiossac2019,Millen2014,Gieseler2014,Rondin2017,Ricci2017,Gieseler2018,Hoang2018,rossi2020experimental}.

\noindent
\section*{Methods}
Here we give some {additional} specifics of the two examples considered in the main text, i.e., the thermal quench of a simple harmonic oscillator and the optical parametric oscillator coupled to a thermal bath. In doing so we follow closely~\cite{Genoni2016} where the example of the optical parametric oscillator is described in details. 
The bath is described, in both our examples, by the initial single mode CM $\sigma_B=(\rm{n}_{\rm{th}}+1/2)\mathbb{I}$. The excitation-exchange interaction Hamiltonian, also common to both the examples, is given by
\begin{equation}
    \hat{H}_{int}=\sqrt{\gamma}\hat{\xb}^T\hat{\xb}_{B}=\frac{1}{2}(\hat{\xb},\hat{\xb}_{B})
    \left(
\begin{array}{c|c}
0 & \mathcal{C} \\
\hline
\mathcal{C} & 0 \\
\end{array}
\right)
  \left(\begin{array}{c}\hat{\xb}_{}\\
  \hat{\xb}_{B}
  \end{array}\right),\,\,\, \mathcal{C}=\begin{pmatrix}
  \sqrt{\gamma} & 0 \\
  0 & \sqrt{\gamma}
  \end{pmatrix}.
\end{equation}
where we have introduced the quadratures of the bath $\hat{\xb}_{B}$. A general-dyne, noisy measurement is described, in the Gaussian formalism, by giving the CM corresponding to the state over which one project. For an ideal general-dyne measurement on the single output mode, the state is pure and the CM is given by the $2\times 2$ matrix
\begin{equation}
2\sigma_{m}=R[\varphi]^T\begin{pmatrix}
  s & 0 \\
  0 & 1/s
  \end{pmatrix}R[\varphi],
\end{equation}
where $R[\varphi]$ is a rotation matrix and $s>0$. Note that, given the form of the excitation-exchange  interaction Hamiltonian, $s=0$ corresponds to homodyne detection of the $x-$quadrature of the output mode and thus the indirect monitoring of the $p-$quadrature of the system, $s=1$ corresponds to heterodyne detection on the output mode, and $s=\infty$ to indirect monitoring of the $x-$quadrature of the system (homodyne detection of the $p-$quadrature of the output mode). In order to account for noisy measurements, the CM $\sigma_m$ needs to be modified by acting on it with the dual of a CP Gaussian map. The reason for this stems from the fact that a noisy measurement can be seen as the action of a CP Gaussian map on the state of the system previous to an ideal general-dyne measurement~\cite{Genoni2016}. For our simple case, the CM $\sigma_m$ for a general-dyne detection with efficiency $\eta\in[0,1]$ and additive Gaussian noise $\Delta$ is given by
\begin{equation}
2\sigma_{m}=R[\varphi]^T\begin{pmatrix}
  s/\eta & 0 \\
  0 & 1/(s\eta)
  \end{pmatrix}R[\varphi]+\begin{pmatrix}
  (1-\eta)/\eta+\Delta & 0 \\
  0 & (1-\eta)/\eta+\Delta
  \end{pmatrix}.
\end{equation}
Given the measurement CM and the interaction Hamiltonian, we can obtain the measurements matrices $\Gamma, C$. For both our examples, these are given by
\begin{align}
    &\Gamma^T=\frac{1}{\sqrt{2}}\Omega\mathcal{C}\sigma_B (\sigma_B+\sigma_m)^{-1/2}\\
    &C^T=-\sqrt{2}\mathcal{C}\Omega (\sigma_B+\sigma_m)^{-1/2}
\end{align}
It should be noted that, in the limit of $\eta\rightarrow 0$ and/or $\Delta\rightarrow\infty$, i.e., for zero efficiency of the detectors and/or infinite additive Gaussian noise, the conditional dynamics converges to the unconditional one. This is intuitive, given that in both these situations no information about the system is acquired by the inefficient/very noisy detection scheme. However, this highlight also another interesting aspect of this type of noisy general-dyne detection: no matter how large the additive noise or how inefficient the detectors, the conditional dynamics will never increase the uncertainty on the state of the system more than the unconditional dynamics.

Finally, we specify the Hamiltonians of the two systems that we consider and we explicit the parameters chosen in Fig.\ref{panels}. For the thermal quench of the harmonic oscillator, we chose a thermal occupation number of the bath such that $2n_{\rm{th}}+1=100$. The Hamiltonian of the system, comprising a quadratic term and a linear drive, is given by
\begin{equation}
    \hat{H}_{s}=\frac{1}{2}\hat{\xb}^T
\begin{pmatrix}
  \omega & 0 \\
  0 & \omega
  \end{pmatrix}\hat{\xb}+\bm{d}^T\Omega\hat{\xb},
\end{equation}
where $d=-(\sqrt{2}\mathcal{E}\cos\theta,\sqrt{2}\mathcal{E}\sin\theta)$ and the driving term correspond, in terms of annihilation and creation operators of the system oscillator, to $\hat{H}_{\rm{drive}}=i\mathcal{E}(\hat{a}e^{i\theta}-\hat{a}^{\dag}e^{-i\theta}).$ Here as in the second example, we work in natural units. Furthermore, we set $\omega=1$ so that all other quantities appearing are in units of $\omega$. In Fig.~\ref{panels} we have chosen $\theta=0, \mathcal{E}=2,$ and $\gamma=1/10$. For the measurement, we chose the rotation matrix to be the identity, i.e., $\varphi=0$, and we consider an ideal measurement $\Delta=0,\,\eta=1$, with $s=\{0,1,\infty\}$. Furthermore, the initial mean values of the oscillator quadratures is $\bar{\mathbf{x}}(t=0)=(1,1)$. The drift and diffusion matrices are readily obtained as
\begin{align}
    &A=\begin{pmatrix}
    -\gamma/2 & \omega \\
    -\omega & -\gamma/2
    \end{pmatrix}\\
    &D=\gamma(n_{\rm{th}}+1/2)\mathbb{I}.
\end{align}

For the optical parametric oscillator, we chose the bath in the vacuum state $n_{\rm{th}}=0$, which is a reasonable assumption for optical modes. The (effective) Hamiltonian of the system is given by
\begin{equation}
    \hat{H}_{s}=\frac{1}{2}\hat{\xb}^T
\begin{pmatrix}
  0 & -\chi \\
  -\chi & 0
  \end{pmatrix}\hat{\xb}.
\end{equation}
The unconditional dynamics of the oscillator is stable only if $\gamma>2\chi$.
The drift and diffusion matrices are given by
\begin{align}
    &A=\begin{pmatrix}
    -\chi-\gamma/2 & 0 \\
    0 & \chi-\gamma/2
    \end{pmatrix}\\
    &D=\gamma(n_{\rm{th}}+1/2)\mathbb{I}.
\end{align}
Note that, as expected, only the reversible part of the drift matrix ($A$) changes with respect to the previous example. In Fig.~\ref{panels}, we set $\chi=1$ so that all other quantities appearing are in units of $\chi$. We chose the coupling constant to be $\gamma=2.001$, i.e., the parametric oscillator is close to the instability point $\gamma=2\chi$. The initial expectation value of the quadratures is chosen in the origin of phase-space. For the measurement, we chose the rotation matrix to be the identity, i.e., $\varphi=0$, and we consider an efficient measurement ($\eta=1$) with a small additive noise $\Delta=0.1$. As before $s=\{0,1,\infty\}$. We add Gaussian noise in such a way to have an unconditional dynamics which, for different detection schemes, leads the system to steady-states with different purities. It should be noted that for a thermal bath at zero temperature and ideal general-dyne measurements, the steady-state of the conditional dynamics would always be a pure state (in contrast to the unconditional steady-state which is always a mixed state).

\noindent
\section*{Data availability Statement.} 
No dataset has been produced for this manuscript.

\section*{Code availability} 
Codes are available upon request from the authors.   

\acknowledgments

The authors would like to thank Mario A Ciampini, Marco Genoni, Nikolai Kiesel, Eric Lutz, Kavan Modi, Albert Schliesser, and Alessio Serafini for stimulating discussions. AB acknowledges the hospitality of the Institute for Theoretical Physics and the ``Nonequilibrium quantum dynamics" group at Universit\"at Stuttgart, where part of this work was carried out. The authors acknowledge financial support from H2020 through the MSCA IF pERFEcTO (Grant Agreement nr. 795782) and  Collaborative Project TEQ (Grant Agreement No.  766900), the Angelo della Riccia Foundation (R.D.~19.7.41.~n.979,~Florence),  the S{\~a}o Paulo Research Foundation (FAPESP) (grant nr. 2018/12813-0 and 2017/50304-7), the DfE-SFI Investigator Programme (Grant No. 15/IA/2864), the Leverhulme Trust Research Project Grant UltraQute (grant nr.~RGP-2018-266), COST Action CA15220, and the Royal Society Wolfson Research Fellowship scheme (RSWF\textbackslash R3\textbackslash183013). GTL and MP are grateful to the SPRINT programme supported by FAPESP and Queen's University Belfast.

\section{Competing Interests}
The authors declare no competing interests.

\section{Author Contributions}
M.P. provided the initial project direction; A.B. and L.M. carried out the core of the calculations and derivation with the input from G.T.L. and M.P.; all authors contributed to the writing up of the manuscript. 

%\section{Additional Information}
%{\bf Supplementary Information} is available for this paper at XXXXXX. 

\end{flushleft}
\bibliography{references2.bib}

\pagebreak
\widetext
\begin{center}
\textbf{\large Supplemental Materials: Entropy Production in Continuously Measured Gaussian Quantum Systems}
\end{center}
%%%%%%%%%% Merge with supplemental materials %%%%%%%%%%
%%%%%%%%%% Prefix a "S" to all equations, figures, tables and reset the counter %%%%%%%%%%
\setcounter{equation}{0}
\setcounter{figure}{0}
\setcounter{table}{0}
\setcounter{page}{1}
\makeatletter
\renewcommand{\theequation}{S\arabic{equation}}
\renewcommand{\thefigure}{S\arabic{figure}}
\renewcommand{\bibnumfmt}[1]{[S#1]}
\renewcommand{\citenumfont}[1]{S#1}

Here we show how to connect $\mathcal{I}$ to the mutual information and how to derive the stochastic entropy production and flux rates from the phase-space dynamics. We furthermore discuss how to obtain the averaged version of these quantities encountered in the main text. % For completeness, we also report the specifics of the two physical examples considered in the main text in order to make all the results reproducible.

\section*{Supplementary Note 1. Mutual Information and Negativity of the informational term}
Here we show that the integrated informational quantity $\mathcal{I}$ is equal to minus the mutual information between the random variable $\mathbf{X}$, representing the position in phase-space, and $\bar{\mathbf{X}}$ described by the stochastic process in Eq.~(3) of the main article. In order to do so, we note that the conditional and unconditional CMs are related by $\sigma_{\rm{uc}}=\sigma+V$, where $V(t)$ is the positive defined CM of the stochastic process $\bar{\mathbf{X}}$. In order to properly characterize $V$, note that for the conditional dynamics the first momenta $\bar{\xb}$ correspond to a classical stochastic process which is in itself Gaussian and, therefore, characterized by its first momenta $\bar{\xb}_{\rm{uc}}=\mathbb{E}[\bar{\xb}]$ (corresponding to the first momenta of the unconditional evolution and evolving according to $d\bar{\xb}_{\rm{uc}}/dt=A\bar{\xb}_{\rm{uc}}+\mathbf{b}$) and the corresponding noise covariance matrix $V$. The evolution of $V$ can be found directly from the Langevin equation in Eq. (3) of the main article and reads 
\begin{equation}
    \frac{dV}{dt} = AV + V A^T + \chi(\sigma).
\end{equation}
Interestingly, we see that the quantum covariance matrix $\sigma(t)$ affects the diffusion coefficient of the noise CM.

Consider the unconditional and conditional Wigner functions $W_{\rm{uc}}=\mathcal{N}(\bar{\xb}_{\rm{uc}},\sigma_{\rm{uc}})$ and $W=\mathcal{N}(\bar{\xb},\sigma)$ (here $\mathcal{N}$ stands for the normal distribution). We can identify the conditional Wigner function with the conditional probability (density) of getting $\bm{X}=\xb$ given $\bar{\bm{X}}=\bar{\xb}$ and write, with a small abuse of notation, $W=p(\xb|\bar{\xb})$. Analogously, we identify $W_{\rm{uc}}=p(\xb)$ in such a way that, given the normal distribution for the stochastic variable $\bar{X}$, we have
\begin{equation}
    p(\xb)=\sum_{\bar{\xb}}p(\bar{\xb})p(\xb|\bar{\xb}),
\end{equation}
which, for Gaussian states, can be shown to be in accordance with the relation $\sigma_{\rm{uc}}=\sigma+V$. Here $p(\bar{\xb})=\mathcal{N}(\bar{\xb}_{\rm{uc}},V)$ is the Gaussian pdf describing the statistics of the \textit{Kalman filtered} outcomes $\bar{\xb}$~\cite{PhysRevLett.94.070405}. Note that, we use sums instead of integrals for the sake of clarity, however the processes that we are considering are always continuous and described properly by probabilities densities.

The quantity $\mathcal{I}$, whose rate $\dot{\mathcal{I}}$ appears in the main text, being the difference between the Wigner entropies of the conditional and unconditional dynamics, can be written as
\begin{equation}
    \mathcal{I}=\sum_{\xb}p(\xb|\bar{\xb})\log p(\xb|\bar{\xb})-H(\bm{X})=H(\bm{X}|\bar{\bm{X}})-H(\bm{X}),
\end{equation}
where we have used the fact that the Wigner entropy of the conditional dynamics is actually \textit{independent} of the outcome $\bar{\xb}$ of the stochastic process, thus $\sum_{\xb}p(\xb|\bar{\xb})\log p(\xb|\bar{\xb})=\sum_{\bar{\xb}}p(\bar{\xb})\sum_{\xb}p(\xb|\bar{\xb})\log p(\xb|\bar{\xb}).$ Finally, we can use the properties of the conditional entropy 
\begin{align}
    & H(A|B)-H(B)=H(B|A)-H(A)\\
    &2 I(A:B)=H(A)+H(B)-H(A|B)-H(B|A),
\end{align}
to conclude that 
\begin{equation}
    \mathcal{I}=-I(\bm{X}:\bar{\bm{X}})\leq 0.
\end{equation}

While we were able to connect $\mathcal{I}$ with a mutual information, the result about its negativity can be readily obtained from the convexity of Wigner entropy. In particular, we have
\begin{equation}
    \mathcal{I}=S(W)-S(W_{\rm{uc}})\leq S(W)-\int d\bar{\mathbf{x}}S(W)p(\bar{\xb})=0,
\end{equation}
where in the last equality we have used again the fact that the Wigner entropy of the conditioned dynamics is a deterministic quantity, i.e., it does not depend on the realization of the stochastic process. We stress once more that this is a peculiarity of the Gaussian formalism. 

Finally, we prove that the inequality is saturated if and only if $\sigma(t)=\sigma_{\rm{uc}}(t)$, i.e., only when the CM of the conditioned dynamics is equal to the one of the unconditional dynamics. To show this, let us consider the definition of $\mathcal{I}$ in terms of the purity, we have
\begin{equation}
    \mathcal{I}=\frac{1}{2}\log\left(\frac{|\sigma|}{|\sigma_{\rm_{uc}}|}\right)=-\frac{1}{2}\log\left(\frac{|\sigma+V|}{|\sigma|}\right).
\end{equation}
Using the properties of the determinant, and the fact that the CM $\sigma$ is positive definite, we can write the argument of the logarithm as
\begin{equation}
  |\sigma+V|\cdot |\sigma|^{-1}=|\mathbb{I}+\sqrt{\sigma}^{-1}V\sqrt{\sigma}^{-1}|.
\end{equation}
From the positivity of $V$, which guarantees that it can be written in the form $LL^{T}$, we deduce that $\sqrt{\sigma}^{-1}V\sqrt{\sigma}^{-1}$ is also positive. Thus, we can write  
\begin{equation}
        \mathcal{I}=-\frac{1}{2}\log\left(|\mathbb{I}+\sqrt{\sigma}^{-1}V\sqrt{\sigma}^{-1}|\right)=-\frac{1}{2}\log\left(\Pi_{i=1,2}(1+\lambda_i)\right),
\end{equation}
where $\lambda_i\geq 0$ are the eigenvalues of the positive definite matrix $\sqrt{\sigma}^{-1}V\sqrt{\sigma}^{-1}$. From this expression it is immediate to conclude that $\mathcal{I}=0$ if and only if $V=0$, i.e., $\sigma=\sigma_{\rm{uc}}$.

{Finally, it should be noted that our informational quantity $\mathcal{I}$ (and its rate) is connected to the Groenewold--Ozawa quantum-classical information (QCI) commonly encountered in the literature~\cite{Sagawa2008}. Indeed, both these quantities are estimators  of the ``information gain''. Loosely speaking, they both represent the amount of information obtained about the system from the measurement. While the QCI compares the original entropy of the system with the final entropy, conditioned on the measurement outcome, the quantity we consider here characterises how much information is gained from reading out the outcomes of the measurements with respect to not doing it. In a simple scheme with a single measurement on a system $A$ with outcome $Z$, the two quantities are formally related by $\mathcal{I}(A':Z)-I_{QC}=S(A')-S(A)$, where $A'$ represents the state of the system after the measurement (see Ref.~\cite{inpreparation} for further details).}

\section*{Supplementary Note 2. Derivation of the stochastic Fokker-Planck equation using Ito-calculus}
Given that, for the conditioned dynamics, the first momenta $\bar{\xb}$ follow a stochastic process, the Wigner function is a function of stochastic variables. Thus, in order to obtain the corresponding Kushner-like dynamical equation in phase-space, we apply Ito-lemma to the Wigner function. The Ito-formula for diffusive processes reads (index summation convention understood)
\begin{align}
    dW&=\bigg(\partial_{\sigma}W\partial_{t}\sigma+\frac{\partial W}{\partial\bar{\xb}_i}(A\bar{\xb}+\bm{b})_i+\frac{1}{2}\frac{\partial^2 W}{\partial\bar{\xb}_i\partial\bar{\xb}_j}\chi(\sigma)_{ij}\bigg) dt+\frac{\partial W}{\partial\bar{\xb}_i}(\sigma C^T+\Gamma^T)_{ij}d\mathbf{w}_j\\
    &=\bigg[\partial_{\sigma}W\partial_{t}\sigma+W\bigg((\xb-\bar{\xb})^T\sigma^{-1}(A\bar{\xb}+\bm{b})+\frac{1}{2}(\xb-\bar{\xb})^T\sigma^{-1}\chi(\sigma)\sigma^{-1}(\xb-\bar{\xb})-\frac{1}{2}{\rm{Tr}}[\sigma^{-1}\chi(\sigma)]\bigg)\bigg] dt\\
    &+W(\xb-\bar{\xb})^T\sigma^{-1}(\sigma C^T+\Gamma^T)d\mathbf{w}.
\end{align}
Using the fact that $D,\chi$ are symmetric, and the properties of the divergence, it is easy to show that the first term on the RHS gives
\begin{align}
    &\partial_{\sigma}W\partial_{t}\sigma=-{\rm{div}}[J_{\rm{a}}]\\
    &J_{\rm{a}}=A (\xb-\bar{\xb}) W - \frac{1}{2} \Bigg( D - \chi(\sigma) \Bigg) \nabla W,
\end{align}
while the other terms give
\begin{align}
&\frac{1}{2}W(\xb-\bar{\xb})^T\sigma^{-1}\chi\sigma^{-1}(\xb-\bar{\xb})-\frac{1}{2}W\,\rm{Tr}[\sigma^{-1}\chi]=-\frac{1}{2}{\rm{div}}(-\chi\nabla W)\\
&W(\xb-\bar{\xb})^T\sigma^{-1}[A\bar{\xb}+\bm{b}]=-{\rm{div}}(W(A\bar{\xb}+\bm{b}))\\
&W(\xb-\bar{\xb})^T\sigma^{-1}(\sigma C^T+\Gamma^T)d\mathbf{w}=-{\rm{div}}(W(\sigma C^T+\Gamma^T)d\mathbf{w})
\end{align}
Putting all the pieces together we finally obtain the stochastic Fokker--Planck equation
\begin{align}\label{StFP}
    dW&=-{\rm{div}}[J_{\rm{a}}+W(A\bar{\xb}+\bm{b})-\frac{1}{2}\chi\nabla W] dt-{\rm{div}}[W(\sigma C^T+\Gamma^T)d\mathbf{w}]\\ \nonumber
    &=-{\rm{div}}[(A \xb+\bm{b})W-\frac{1}{2}D\nabla W] dt-{\rm{div}}[W(\sigma C^T+\Gamma^T)d\mathbf{w}]\\ \nonumber
    &=-{\rm{div}}[J dt+J_{\rm{sto}}],
\end{align}
where $J=(A \xb+\bm{b})W -\frac{1}{2}D\nabla W $ and $J_{\rm{sto}}=W(\sigma C^T+\Gamma^T)d\mathbf{w}$. It is interesting to note that, the deterministic current coincides (in form) with the one we would have for the unconditional dynamics. 

\section*{Supplementary Note 3. Stochastic Entropy Flux and Production rates}
Let us consider the case of conditional dynamics, in which the evolution of the system in phase-space is described by the stochastic Kusher-like equation~\eqref{StFP}. Note that, while the Wigner entropy rate is a deterministic quantity (since it depends on the deterministic dynamics of the CM), the Wigner function is not. Thus, in order to work out the Wigner entropy rate we need to resort to Ito-lemma. It is then easy to see that the entropy increment 
\begin{align}
    dS&=-\int d\xb d(W\log W)\\
    &=-\int d\xb\bigg[\bigg(\frac{\partial W\log W}{\partial\sigma}\frac{\partial\sigma}{\partial t}+\frac{\partial W\log W}{\partial\bar{\xb}_i}(A\bar{\xb}+\bm{b})_i+\frac{1}{2}\frac{\partial^2 W\log W}{\partial\bar{\xb}_i\partial\bar{\xb}_j}\chi_{ij}\bigg)dt\\
    &+\frac{\partial W\log W}{\partial\bar{\xb}_i}(\sigma C^T+\Gamma^T)_{ij} d\mathbf{w}_j\bigg],
\end{align}
results in
\begin{align}
    dS&=\int d\xb\, {\rm{div}}\bigg[\bigg(A\xb W+\bm{b} W-\frac{1}{2}(D-\chi)\nabla W\bigg)dt+(\sigma C^T+\Gamma^T)d\mathbf{w} W\bigg]\log W\\
    &=\int d\xb\, {\rm{div}}\bigg[(J+J_2)dt+J_{\rm{sto}}\bigg]\log W,
\end{align}
where we have introduced $J_{2}=\frac{1}{2}\chi\nabla W$, while $J$ and $J_{\rm{sto}}$ are the ones defined in the main text. It is interesting to note that the term containing the second derivatives of $W$ with respect to the first momenta vanishes identically. 

Note that, in analogy with the unconditional dynamics case~\cite{Santos2018}, the Hamiltonian part of the current, i.e., $J_{\rm{rev}}$, does not contribute to the entropy rate. Indeed, such a term is divergence-less and can be dropped out. The same is true also for the $J_{\rm{sto}}$ part of the current, but not for $J_{2}$ which encodes the measurement matrices. Thus, integrating by parts, we remain with
\begin{align}
    dS=-\int \frac{d\xb}{W}(J_{\rm{irr}}+J_2)^T \nabla W\,dt
\end{align}
Expressing the gradient of the Wigner function in terms of $J_{\rm{\rm{irr}}}$ and $J_{2}$, the entropy increment can be cast in the form
\begin{align}\label{dS}
    dS&=2\int \frac{d\xb}{W}J_{\rm{\rm{irr}}}^TD^{-1}J_{\rm{\rm{irr}}}dt -2\int \frac{d\xb}{W}J_{2}^T\chi^{-1}J_{2}dt -2\int d\xb J^{T}_{\rm{irr}}D^{-1}A_{\rm{irr}}\xb dt.
\end{align}
We identify the last terms as the entropy flux increment and the first two terms as the entropy production one. In the unconditional dynamics case, only the first term is present for the entropy production and, in the literature, such term is identified with the entropy production (since it is a positive semi-definite \textit{quadratic} form of the currents). In the conditional case, we see that on top of the positive semi-definite quadratic form, the second term quadratic in the currents is negative semi-definite due to the fact that $\chi$ is a positive semi-definite matrix by construction. Nonetheless, we incorporate this term in the entropy production and recognize in the last term the entropy flux with the same form as in the unconditional case. Thus, we end up with the following expressions for the entropy flux and production in terms of the parameters of the model
\begin{align}
    & \frac{dS}{dt}=\frac{1}{2}{\rm{Tr}}[2A+\sigma^{-1}(D-\chi)],\\
    & d\phi_{\bar{\xb}}=-{\rm{Tr}}[A_{\rm{irr}}]dt-2{\rm{Tr}}[A_{\rm{irr}}^TD^{-1}A_{\rm{irr}}\sigma]dt-2\bar{\xb}^TA_{\rm{irr}}^TD^{-1}A_{\rm{irr}}\bar{\xb}dt\\
    & d\pi_{\bar{\xb}}=2{\rm{Tr}}[A_{\rm{irr}}]dt+2{\rm{Tr}}[A_{\rm{irr}}^TD^{-1}A_{\rm{irr}}\sigma]dt+2\bar{\xb}^TA_{\rm{irr}}^TD^{-1}A_{\rm{irr}}\bar{\xb}dt+\frac{1}{2}\rm{Tr}[\sigma^{-1}(D-\chi)].
\end{align}
Note that these quantities still depends on the stochastic variable $\bar{\xb}$. In order to obtain the entropy flux and production rates, in which we are interested, we need to average these expressions over all possible trajectories of the system. Given the Gaussian nature of the problem, the average is trivial to perform using the Gaussian pdf
\begin{equation}
    p(\bar{\xb},t;\bar{\xb}_{\rm{uc}},V)=\frac{1}{(2\pi)^n\sqrt{\det V}}e^{-\frac{1}{2}(\bar{\xb}-\bar{\xb}_{\rm{uc}})^T V^{-1}(\bar{\xb}-\bar{\xb}_{\rm{uc}})},
\end{equation}
where $V$ is the noise matrix we introduced above. The result of the average is
\begin{align}
    & \mathbb{E}\left[\frac{d\phi_{\bar{\xb}}}{dt}\right]=-{\rm{Tr}}[A_{\rm{irr}}]-2{\rm{Tr}}[A_{\rm{irr}}^TD^{-1}A_{\rm{irr}}\sigma_{\rm{uc}}]-2\bar{\xb}_{\rm{uc}}^TA_{\rm{irr}}^TD^{-1}A_{\rm{irr}}\bar{\xb}_{\rm{uc}} =\Phi_{\rm{uc}}\\
    & \mathbb{E}\left[\frac{d\pi_{\bar{\xb}}}{dt}\right]=2{\rm{Tr}}[A_{\rm{irr}}]+2{\rm{Tr}}[A_{\rm{irr}}^TD^{-1}A_{\rm{irr}}\sigma_{\rm{uc}}]+2\bar{\xb}_{\rm{uc}}^TA_{\rm{irr}}^TD^{-1}A_{\rm{irr}}\bar{\xb}_{\rm{uc}}+\frac{1}{2}\rm{Tr}[\sigma_{\rm{uc}}^{-1}D]\\ \nonumber
    &+\frac{1}{2}{\rm{Tr}}[(\sigma^{-1}-\sigma_{\rm{uc}}^{-1})D-\sigma^{-1}\chi]=\Pi_{\rm{uc}}+\dot{\mathcal{I}},
\end{align}
where we have used the explicit expressions for the entropy flux ($\Phi_{\rm{uc}}$) and production ($\Pi_{\rm{uc}}$) rates for the unconditional dynamics given by
\begin{align*}
    & \Phi_{\rm{uc}} =-2\int d^{2n}\xb J^{T}_{\rm{irr}}D^{-1}A_{\rm{irr}}\xb=-{\rm{Tr}}[A_{\rm{irr}}]-2{\rm{Tr}}[A_{\rm{irr}}^T D^{-1}A_{\rm{irr}}\sigma_{\rm{uc}}]-2\bar{\xb}_{\rm{uc}}^{T}A_{\rm{irr}}^T D^{-1}A_{\rm{irr}}\bar{\xb}_{\rm{uc}}\\
    & \Pi_{\rm{uc}}=2\int \frac{d^{2n}\xb}{W_{\rm{uc}}}J_{\rm{irr}}^TD^{-1}J_{\rm{irr}}=2{\rm{Tr}}[A_{\rm{irr}}]+2{\rm{Tr}}[A_{\rm{irr}}^T D^{-1}A_{\rm{irr}}\sigma_{\rm{uc}}]+\frac{1}{2}Tr[\sigma_{\rm{uc}}^{-1}D]+2\bar{\xb}_{\rm{uc}}^{T}A_{\rm{irr}}^T D^{-1}A_{\rm{irr}}\bar{\xb}_{\rm{uc}},
\end{align*}

%\textcolor{red}{
%\begin{equation}
%\Phi_{\rm{uc}} =-2\int d^{2n}\xb J^{T}_{\rm{irr}}D^{-1}A_{\rm{irr}}\xb,\quad \Pi_{\rm{uc}}=2\int \frac{d^{2n}\xb}{W_{\rm{uc}}}J_{\rm{irr}}^TD^{-1}J_{\rm{irr}}.
%\end{equation}
%In the literature
%}
We see that the stochastic quantity $d\Phi_{\bar{\xb}}$ averages exactly to the unconditional entropy flux. This result strengthen our assumption in the main text about the linearity of the stochastic flux with respect to the state of the system. Indeed, we have shown here how to arrive at an expression for the stochastic flux which is linear in the state, in terms of the dynamics of the Wigner function in phase-space. The crucial step is to identify the entropy production rate as the term quadratic in the currents. While this is less direct in this case compared to the unconditional dynamics, in which the quadratic term is positive definite, it squares well with the presence of the current $J_2$ which is related to the measurement and can, in principle, make the entropy production rate negative.

%\section*{Supplementary References}

%\bibliography{references2.bib}

\end{document}